\documentclass{aastex}                
\usepackage{spr-astr-addons}    
\usepackage{url}

\begin{document}

\title{Minimal spanning tree algorithm for $\gamma$-ray source detection in sparse photon images:
cluster parameters and selection strategies.} 

\shorttitle{The Minimal Spanning Tree algorithm for $\gamma$-ray source detection}
\shortauthors{R. Campana et al.}

\author{R.~Campana\altaffilmark{}}
\affil{INAF/IAPS, Rome and INAF/IASF-Bologna, Bologna, Italy.} 
\and 
\author{E.~Bernieri}
\affil{INFN/LNF and Department of Physics, University of Roma Tre, Rome, Italy.}
\and
 \author{E.~Massaro}
\affil{Department of Physics, University of Rome ``La Sapienza'', Rome, Italy.}
\and 
\author{F.~Tinebra}
\affil{Department of Physics, University of Rome ``La Sapienza'', Rome, Italy.}
\and 
\author{G.~Tosti}
\affil{Department of Physics, University of Perugia, Perugia, Italy.}
\email{riccardo.campana@inaf.it}

\begin{abstract}
The minimal spanning tree (MST) algorithm is a graph-theoretical cluster-finding method.
We  previously applied it to $\gamma$-ray bidimensional images, showing that it is quite sensitive in finding 
faint sources.
Possible sources are associated with the regions where the photon arrival directions clusterize. 
MST selects clusters starting from a particular ``tree" connecting all the point of the image 
and performing a cut based on the angular distance between photons, with a number of events higher 
than a given threshold.
In this paper, we show how a further filtering, based on some parameters linked to the cluster 
properties, can be applied to reduce spurious detections.
We find that the most efficient parameter for this secondary selection is  the
\emph{magnitude} $M$ of a cluster, defined as the product of its number of events by
its \emph{clustering degree}.  
We test the sensitivity of the method by means of simulated and real \textit{Fermi}-Large Area Telescope (LAT)
fields. 
Our results show that $\sqrt{M}$ is strongly correlated with other statistical significance parameters, 
derived from a wavelet based algorithm and maximum likelihood (ML) analysis, and that it can be used as 
a good estimator of statistical significance of MST detections.
We apply the method to a 2-year LAT image at energies higher than 3 GeV, and we show the presence of
new clusters, likely associated with BL Lac objects.
\end{abstract}

  
\section{Introduction}\label{s:introduction}

Telescopes for medium and high energy $\gamma$-ray astronomy usually detect individual photons 
by means of the electron-positron pairs generated through the instrument:
from their trajectories it is possible to reconstruct the original direction of
the photon, with an uncertainty that decreases from a few degrees below 100 MeV to a
few tenths of a degree above 1 GeV.
The resulting product is an image where each photon is associated with a direction 
in the sky: discrete sources correspond thus to small size regions with a localized 
concentration of photons higher than in the surroundings.
When the size of this region and the photon spatial distribution are consistent with 
the instrumental point spread function (PSF) the source is considered as point-like, 
otherwise it can be an extended feature or an unresolved group of close sources. 
Various algorithms are applied to the detection of point-like or extended sources: 
among them, Maximum Likelihood \citep{mattox96}, Multi-Resolution Filter \citep{starck95}, 
Multi-Scale Variance Stabilizing Transform \citep{schmitt12}, Optimal Filter \citep{tegmark98}, 
Aperture Photometry \citep{harnden84}, Wavelet Transform analysis \citep{damiani97}, etc.
Some of these (like Wavelet Transform or Optimal Filter) are based on deconvolution 
techniques of the PSF, while other methods search for clusters in the arrival directions 
of photon that, if statistically significant, are considered an indication for the presence of a source.

In previous papers \citep{campana08,massaro09a}, we 
described the \emph{minimal spanning tree} (MST) algorithm for the search of photon clusters
to be associated with $\gamma$-ray source candidates.
This technique has its root in graph theory, and highlights the \emph{topometric} 
pattern of ``connectedness'' of the detected photons.

\begin{figure*}
\centering
\includegraphics[scale=0.42]{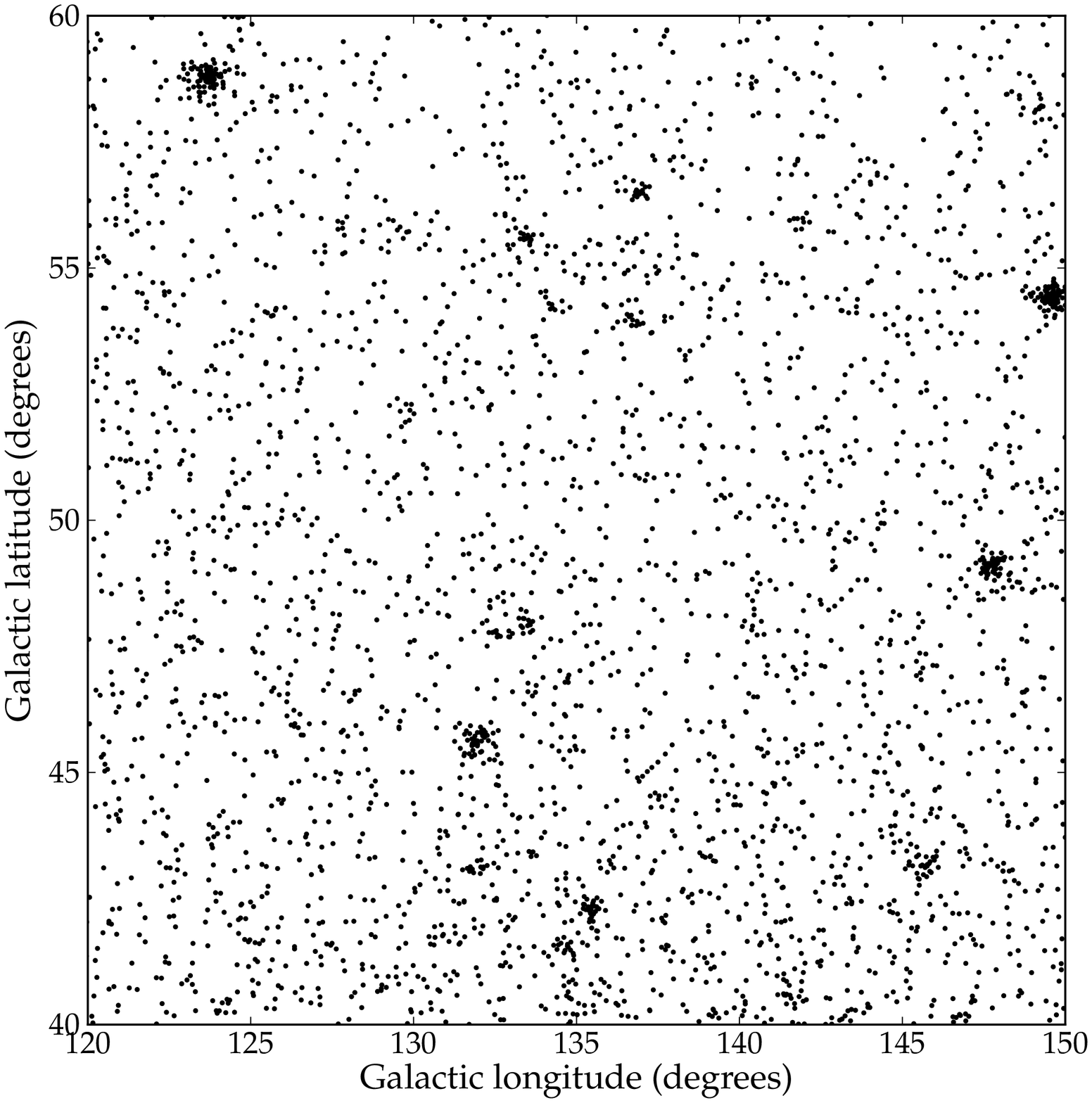}\includegraphics[scale=0.42]{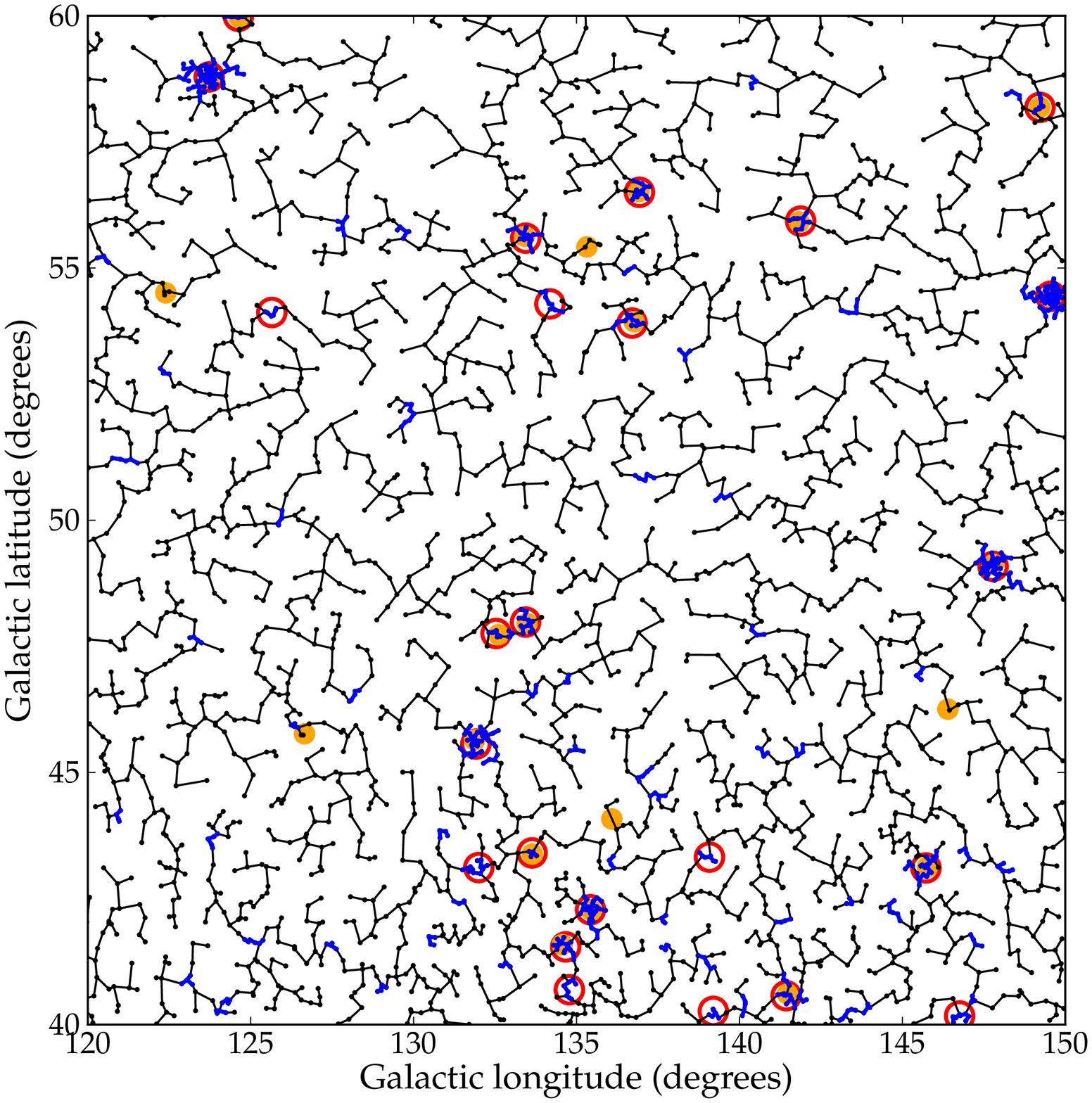}
\caption{A example of MST source detection in a LAT field. Left panel: A region 
($30^{\circ} \times 20^{\circ}$) of the 2-years LAT sky at energies $>$3 GeV centered 
at $l=135^{\circ}$, $b=50^{\circ}$. 
Right panel: the MST between the photons and the clusters (in blue) found applying the 
selection criteria $\Lambda_{\mathrm{cut}} = 0.7\,\Lambda_\mathrm{m}$ and $N_{\mathrm{cut}} = 3$. 
Red circles mark the candidate $\gamma$-ray sources after the secondary selection on $M$ 
(see Sect. \ref{s:clusterparameters}); 
yellow circles mark the positions of sources in the LAT 2FGL catalog, that includes also sources detected only at lower energies.}
\label{fig:mstexample}
\end{figure*}

In the field of high energy astrophysics, source detection with MST was first proposed by  
\cite{digesu83} for the COS-B mission. 
Further developments were made by
\cite{digesu86,debiase86,maccarone86,buccheri88}, all of these before the observations
of \textit{COMPTON/GRO}-EGRET and \textit{Fermi}-Large Area Telescope (LAT).
In other astronomical contexts, MST-based methods were applied to the goal of finding 
galaxy clusters, both in 2 and 3-dimensional surveys and simulations 
\citep{barrow85, bhavsarling88a, bhavsarling88b, plionis92, krzewina96}. 
These authors also show the capabilites of the MST as a filament-finding algorithm.
More recently, MST has been used for studying the clustering of stars by \cite{koenig08} and \cite{schmeja11}, 
who applied some parameters proposed by us \citep{massaro09a} for the cluster selection.
The advantage of MST, and of other cluster-finding algorithms such as DBSCAN \citep{tramacere13}, 
is the capability to quickly find potential $\gamma$-ray sources by examining only the incoming directions of the 
photons, regardless of their energy distribution. This can be useful to find transient sources, that can 
only be observed by performing a periodic and systematic data analysis 
over time intervals of weeks or months, or to produce lists of candidate sources that can be further 
analysed with other well-recognized statistical methods, such as maximum likelihood (ML).

Despite the fact that MST is not the optimal cluster-finding algorithm from the point of view 
of computational speed, it has the great advantage of providing some useful parameters, that are evaluated 
over the entire region under consideration, and are relevant for the selection of
interesting clusters, as we will show in this paper.    
We already applied the MST algorithm for detecting clusters of photons in the 
 $\gamma$-ray sky in order to obtain lists of candidate sources for the 1FGL
and 2FGL Fermi-LAT catalogues \citep{abdo10,nolan12}. 
In this application of MST to LAT data, however, we faced some challenges, 
such as the assessment of the optimal detection threshold, the accuracy of 
sources' coordinates --- relevant for a safe identification of possible counterparts 
of $\gamma$-ray sources --- and the detection significance.

We are mainly interested in the detection of clusters having a rather small number of photons, 
typically less than 10, that could be associated with faint $\gamma$-ray sources. 
This problem is important in $\gamma$-ray astrophysics because many sources, particularly blazars, are highly variable with short duty-cycles, 
and therefore their detection becomes harder with long observations when the integrated background level increases with time and the source signal may fade.
In this paper we present some new methods to improve the 
capabilities of MST for detecting $\gamma$-ray source candidates, based on our experience in 
the analysis of LAT data.
We will show that it is possible to establish some rather simple criteria for the selection
of the best source candidates, applying the method to simulated and real \textit{Fermi}-LAT sky test fields.
We also compare MST results with those obtained applying the Perugia Wavelet (PGW) transform code 
\citep{ciprini07}.
Based on this comparison, we will show that it is possible to obtain a satisfactory estimate of
the statistical detection significance, considering only one of the ``cluster parameters" 
we introduce.

This paper is organised as follows. In Section \ref{s:mstreview} a brief review of the MST method 
is given, and in Section \ref{s:mstapplication} we discuss some problems arising in the 
application of the MST to $\gamma$-ray data. 
In Section \ref{s:localization} we present a way to improve the source localization accuracy, in 
Section \ref{s:clusterparameters} we introduce some parameters able to discriminate between 
good and spurious sources, and in Section \ref{s:testfield} we show the results of the method 
on simulated fields. 
In Section \ref{s:pgw} we compare MST and PGW results, in Section \ref{s:LAT} we analyze 
the true 2-year LAT field and in Section \ref{s:newsources} we discuss 
the properties of new clusters detected by MST in a real 2-year \textit{Fermi}-LAT field and their possible
counterparts. Finally, in Section \ref{s:conclusions} we draw our conclusions.

\section{The MST algorithm for cluster search}\label{s:mstreview}

In the following we outline the major steps in MST source detection;
for a more complete description of the method and of its statistical properties see \cite{campana08}.

Once having a set of $N$ spatially distributed 
\emph{nodes}, one can compute the set $\{\lambda_i > 0\}$ of weighted \emph{edges} 
connecting them: 
the MST is the tree satisfying the condition $\min[\Sigma_i \lambda_i$] \citep{zahn71}.
For a set of points in a Cartesian frame the edges are the lines joining the nodes, 
weighted by their length; for a set of points over a spherical surface, like in 
astronomical imaging, the weights are the angular distances.
Several algorithms for the MST computation are available. 
One widely used was developed by \cite{prim57}: it starts from an arbitrary selected node, finds 
the nearest neighbour and connects them with an edge, which is thus the first MST edge. 
Then it finds the point that is the nearest to any point that is already connected in 
the tree. 
After $N-1$ iterations, the complete MST is found. 
Faster and computationally optimized codes can be found using other theoretical 
properties of the MST, like being a subset of the Delaunay triangulation of the graph.
In particular, we used a fast code for the MST computation that uses the freely available  
\textsc{Boost}\footnote{\url{http://www.boost.org}} and 
CGAL\footnote{\url{http://www.cgal.org}}  libraries.

To extract \emph{only} the photons that cluster in regions with a density higher
than the mean one in the field, i.e. the candidate sources, the following operations are 
performed:

\begin{itemize}

\item {\bf Separation:} remove all the edges having a length 
$\lambda > \Lambda_{\mathrm{cut}}$, 
the \emph{separation value}, that can be defined either in absolute units or in units of the 
mean edge length in the MST: $\Lambda_{\mathrm{m}} = (\Sigma_i \lambda_i)/(N - 1) $; we thus obtain 
a set of disconnected sub-trees.

\item {\bf Elimination:} remove all the sub-trees having a number of nodes 
$ n \leq N_{\mathrm{cut}}$.
We thus remove small random clusters of photons, leaving only the clusters having a size 
greater than the selected lower limit.

\end{itemize}

After the application of these two filters, the remaining set of sub-trees $\{S_k\}$ provides 
a list of clusters, which in our case are candidate $\gamma$-ray sources.

When applied to the $\gamma$-ray sky, we start with a LAT photon list that contains the 
arrival directions of all accepted events, time, energy, and other useful parameters.
In a selected region, typically large enough to cover at least tens of square degrees, 
we consider the photon arrival directions as the nodes in the bi-dimensional graph, 
the edge weight being the angular distance between them, and compute the MST. 
An example of a sky image, at energies higher than 3 GeV, is shown in the left panel of 
Fig. \ref{fig:mstexample}, while the MST that connects the event positions is presented 
in the right panel, where the residual subtrees after the two aforementioned operations 
are marked in blue.

\cite{campana08} studied the distribution of edge lengths in MST \citep{barrow85} for random 
Poissonian fields\footnote{The formal definition of a spatial Poisson process with uniform density involves 
the fact that in each closed subregion the number of points follow a Poisson distribution, and 
the various closed disjoint regions are independent from each other. See, e.g., \cite{diggle03}
for details.}
and found that it is well fitted by a Rayleigh distribution modified by a Fermi-Dirac factor.
When clusters are present the distribution changes and becomes more asymmetric, with an 
excess of both short distances (due to the clusters associated to the sources) and long 
distances (due to the decrease of the mean distance). 
A theoretical derivation of the Rayleigh distribution, for a pure Poissonian field,
can be found in \cite{tinebra11}.
A simple indicator of clusterization is the mean value of the MST length. 
The total length of a MST in a field with a uniformly random 
distribution of nodes is proportional to $\sqrt{ A\,N}$ \citep{gilbert65} where $A$ is the field area. 
The mean length for a random-field MST can be written as:
\begin{equation}
\Lambda_{\mathrm{m}} = C \sqrt{\frac{A}{N}}
\end{equation}
A theoretical upper limit to $C$ equal to $2^{-1/2} \simeq 0.70$ was established by 
\cite{gilbert65}, and our Monte Carlo simulations \citep{campana08} consistently found a $C$ value 
close to 0.65.
Thus, if the mean length for a field is quite lower than this value, it 
can be considered as an indicator of non-random clusterization, i.e. of the presence 
of sources.

\section{MST application to the $\gamma$-ray sky}\label{s:mstapplication}

In the practical applications of MST-based cluster extraction to $\gamma$-ray sky images,
we have to take into account several complications, arising from the 
presence of large scale structures mainly due to the Milky Way high energy emission. 
After excluding a belt along the Galactic equator having a latitude width of the order 
of 15$^\circ$--20$^\circ$, there is still an appreciable decreasing gradient of photon density in 
the directions of the Galactic poles, even at energies higher than 1 GeV. 
Moreover, spatial features extending to high latitudes are apparent above and below the 
Galactic centre and in other parts of the disk.

An important consequence of this inhomogeneous photon density for the MST cluster finding 
is that the mean edge length $\Lambda_{\mathrm{m}}$ does not correspond to the \emph{local} 
one, that is shorter in the high density zones and longer in the low density ones.
This implies that when $\Lambda_{\mathrm{cut}}$ is given as a a fraction of $\Lambda_{\mathrm{m}}$,
it is possible that several spurious clusters are found in the high density zone.
To reduce this effect one can divide the region in a suitable number of overlapping 
subregions in which the photon density has only rather small variations:
clusters are then selected from the central zone of each subregion, avoiding the
boundaries where the clustering can be affected by the discontinuity in the photon
field.
Another problem arises if a bright source is in one of the subregions: the mean edge length 
in the subregion could be quite lower than in the others, affecting thus the  
selection of poorer clusters.
These peculiar subregions require a particular analysis, possibly with different choices
of the boundaries.  
  
However, once the MST has been obtained for a photon field, a good strategy for selecting spatial 
clusters and, among them, those having a high probability to correspond to genuine $\gamma$-ray 
sources, is to apply the following two steps:

\begin{itemize}
\item {\bf Primary selection}: here we use only the two parameters $N_{\mathrm{cut}}$ and 
$\Lambda_{\mathrm{cut}}$ (Sect. \ref{s:mstreview}) to select all the clusters having the minimum 
requirements to be a structure in the field.
As will be shown in Sect. \ref{s:testfieldresults}, it is
usually applied a rather low $N_{\mathrm{cut}}$, and values of $\Lambda_{\mathrm{cut}}$ 
in the range 0.6--0.8 in units of $\Lambda_{\mathrm{m}}$ are shown to minimize the number of spurious clusters:
a value of $\Lambda_{\mathrm{cut}}$ closer to unity would increase the probability of 
selecting clusters due to random fluctuations of the background, while a lower value would
reduce the selection only to very dense clusters.
When searching for point sources one has to take into account the instrumental 
angular resolution, and values of $\Lambda_{\mathrm{cut}}$ corresponding to edge lengths larger than the radius 
of the PSF must be rejected. 
This can be the case for fields with a low mean photon density. 
Moreover, to avoid a selection of clusters not matching the expected point source characteristics, 
a lower threshold value on the cut length should be imposed, corresponding to an angular separation 
a few tenths of a percent greater than the effective PSF radius.
\end{itemize}

After the application of this step, usually we obtain a rather long list of clusters for which 
we compute several quantities, the \emph{cluster parameters}, convenient for the further
selection.
These parameters, defined in Section \ref{s:clusterparameters}, give a measure of the compactness 
of the clusters related to some properties of the field and can be used to take into account the 
expected properties of the $\gamma$-ray source population, such as the angular extension to be 
compared with the instrumental PSF.
Then, a further selection can be performed:

\begin{itemize}
\item {\bf Secondary selection}: it is based on the definition of selection thresholds to be 
applied to the cluster parameters and extracts from the first list those having properties 
expected for best candidates of ``true'' point-like sources. 
The main objective is to maintain in the final list clusters having a high probability to be
associated with genuine astrophysical sources and to eliminate those originating from density 
fluctuations of the cosmic background and residual cosmic rays. 
\end{itemize}

\section{Source location accuracy}\label{s:localization}

A good estimate of source positions in the sky is very important for the secondary
selection, for the comparison between MST results with those of other methods, and for the 
association of possible counterparts.
The simplest approach is the calculation of the centroid position, i.e. the arithmetic 
mean of the coordinates of all photons belonging to a selected cluster.
Generally, this elementary method gives a good estimate of the source coordinates. 
In some cases, however, it can fail to provide a satisfactory location for different reasons: 
$i$) the cluster is in a relatively high background region and its shape can be highly irregular;
$ii$) the separation value is not small enough and the cluster extends in a particular direction 
with a branch well outside the PSF size; 
$iii$) connection and/or proximity with another cluster, producing an enlarged structure; 
$iv$) small number of nodes; 
$v$) sufficient number of nodes but moderate clustering, particularly relevant for sources
with soft spectra.

Starting from the first centroid position defined above, one can refine it by means of:
$i$) using suitable ``weights'' in averaging nodes' coordinates, $ii$) applying a 
further selection of the nodes to be used in the calculation based on their distance from 
the first centroid.
The method of weighted averages is justified by the energy dependence of the PSF radius, quite 
smaller at higher than lower energies.
Weights, therefore, could be related to this additional information, because high energy photons
are expected to have a higher concentration around the `true' source position.
Another possibility, which takes into account this property implicitly, is the use of photons' 
coordinates to establish the weights.
We thus introduced as a weight the $q$-th power of the distance from any node to the nearest 
one, $1/\lambda^q$. 
However, the possible occurrence in a cluster of a couple of events with a very small angular 
separation between them, but located far from the first centroid, would produce a large 
difference of their weighted positions with respect to the initial one.
This effect was found not negligible even for clusters having a particularly high number of nodes
and can be avoided by excluding from the calculation a fraction of nodes at high distance from the 
centroid.

To establish the best values of $q$ and of the fraction of excluded photons \cite{tinebra11} tested the method
by analysing the differences of coordinates with respect to a sample of 130 sources from the 1FGL 
catalogue, selected with different number of photons and cluster parameters, with the exclusion of 
bright sources for which the centroid estimates are generally very good.

It was found that the most convenient solution, also from a computational time 
point of view, was to consider only the fraction of 6/7 of the nodes closest to the initial centroid 
position, together with an exponent $q=1$ (for more details, see Tinebra 2011). 
This choice was a trade-off between avoiding the elimination of events in poor clusters (i.e. with $n < 6$) 
and to  select the fraction of nodes close to the first centroid location in the ones with a high event 
number, which have a high probability of contamination from the surrounding background.
An improvement of coordinates for the 80\% of sources was obtained, while all 
the worsenings were smaller than 0.1 degree, well inside the LAT 
PSF radius\footnote{\url{http://www.slac.stanford.edu/exp/glast/groups/canda/lat_Performance.htm}}, and thus without 
practical effects for the search of counterparts. 

We further verified this method for evaluating the central position of clusters by comparing our MST results, at energies higher than 3 GeV, with
the coordinates of a large sample of 400 active galactic nuclei extracted from the 2 year Fermi-LAT AGN Catalog
2LAC \citep{ackermann11} that reports the most likely radio and optical  counterparts 
of 2FGL sources.
About 90\% of the MST positions have an angular separation smaller than 6$'$.7 ($\sim$0$^\circ$.1) with respect to the source position.
The same fraction of 2FGL positions (estimated by ML) have an angular separation smaller than 4$'$.7 ($\sim$0$^\circ$.08).
On average, the positional accuracy of MST differs with respect to ML by 1$'$.2 (0$^\circ$.02), much lower than the PSF size at these energies.
We can conclude that our algorithm provides an estimate of the clusters' centres good enough to search for reliable counterparts.

\section{Cluster parameters}\label{s:clusterparameters}

In  \cite{campana08} we defined two quantities useful to evaluate the 
``goodness'' of the clusters after the primary selection: 
the number of nodes $n_k$ in the $k$-th cluster and the \emph{clustering degree} 
$g_k = \Lambda_\mathrm{m}/ \lambda_{\mathrm{m},k}$, where $\lambda_{\mathrm{m},k}$ is the mean of 
the edge lengths in the $k$-th cluster.
From a practical point of view, the former parameter increases with the mean flux of the source,
while the latter depends on the spectrum, because sources with a small mean separation between 
nodes usually have a higher fraction of high energy photons.
A secondary selection that uses a severe cut on $g$, particularly in regions of non uniform or 
high background, would reduce the efficiency for the detection of soft sources even with a 
relatively high number of nodes.
To limit this effect it is useful to consider 
the quantity
\begin{equation}\label{eq:magnitude}
M_k =  n_k g_k \, ,
\end{equation} 
that combines the number of nodes with their clustering degree and that we named \emph{magnitude} 
\citep{massaro09a}.
A cluster with a small number of nodes, but with very short edge lengths, can have a magnitude 
comparable to that of a richer but not so dense cluster. 

It is possible to define other similar parameters. 
After the separation step, two sets of edges are obtained: one includes all the edges between 
connected nodes in clusters and the other contains all the edges pertaining to disconnected nodes. 
This latter set represents the background in the field. 
One can then compute the mean values of these sets, $\Lambda^\mathrm{C}_\mathrm{m}$ and $\Lambda^\mathrm{B}_\mathrm{m}$
respectively, and evaluate the corresponding clustering degrees:
$g^\mathrm{C}_k = \Lambda^C_\mathrm{m} / \lambda_{\mathrm{m},k}$ and 
$g^\mathrm{B}_k = \Lambda^\mathrm{B}_\mathrm{m} / \lambda_{\mathrm{m},k}$.
Clearly, in a given field, one has $g^\mathrm{B}_k > g_k > g^\mathrm{C}_k$, 
since on average the ``background'' edges ($\Lambda^\mathrm{B}_\mathrm{m}$) are longer than the ``cluster'' edges ($\Lambda^\mathrm{C}_\mathrm{m}$).
In other words, $g^\mathrm{B}_k$ and $g^\mathrm{C}_k$ quantify the amount of clustering with respect to the background and other sources, respectively.
In fact, $g^\mathrm{B}_k$ and $g^\mathrm{C}_k$ are useful when comparing clusters detected in 
different regions (or subregions).
For each cluster one can also consider the magnitudes $M^B_k$ and $M^C_k$, 
defined in a similar way to $M_k$, but using the corresponding clustering degrees.
In the following applications, we will consider only the magnitude given by Eq. \ref{eq:magnitude} 
and analyse its statistical meaning.

When the primary selection is performed using a too small value of $\Lambda_{\mathrm{cut}}$, it
is possible that a rich cluster of nodes is divided into some minor structures, while it would be 
a unique cluster by applying a slightly larger separation value.
These \emph{satellite clusters}, usually located in the surroundings of a true bright source, 
are generally artifacts of the primary selection and must be properly taken into account
in the subsequent analysis.
For each cluster, we can introduce a new quantity, named the \emph{proximity value} $p$, equal to
the angular distance to the nearest cluster.
Thus, clusters with $p$ higher than a few times the PSF radius (e.g. $\sim 1^{\circ}$ in LAT images at
GeV energies) have a high probability to be genuine sources, while $p < 0^{\circ}.5$ can be the 
indication for a satellite cluster.
The spatial structures of all the clusters with small $p$ needs further investigations before taking
a final decision on them.

Two other parameters that can provide useful information on the cluster extension are the
\emph{cluster radius} $R_\mathrm{c}$ and the \emph{median radius} $R_\mathrm{m}$, defined as the 
radius of the circle centered at the improved centroid of the cluster (see Sect. \ref{s:localization}) 
and which includes all and half of the $n_k$ nodes, respectively.
A high value of clustering parameter $g_k$ usually corresponds to small values of these radii,
although in a few cases this property is not verified.
This can occur when the cluster exhibits an elongated feature in some direction, likely
due to some background pattern.
Again, clusters with $R_\mathrm{c}$ or $R_\mathrm{m}$ values higher than those expected for a radially 
symmetric structure, typically comparable or lower than the effective radius of the instrumental PSF, 
should be further investigated to verify whether they can be associated with genuine sources or not.

We found that the magnitude is the most efficient parameter for the secondary selection.
In Sect. \ref{s:testfield}, we will show that by applying a suitable threshold to the $M$ value, 
the majority of spurious clusters, i.e. not originated by true sources, can be rejected with a 
limited loss of good ones.

\section{MST application to test fields}\label{s:testfield}

\subsection{Description of the simulated test fields}

To verify the capability of the MST method and to devise the criteria for the secondary selection,
in order to provide a high efficiency in accepting clusters associated with ``true'' sources and
rejecting the spurious ones, we applied the above procedure to simulated photon fields with 
properties similar to the LAT $\gamma$-ray sky outside the Galactic belt.
To work well with the MST a low density field is preferred: in fact, when the mean angular 
distance between the photons is smaller than the angular precision of the arrival direction,
sources can be close to the confusion limit, and the detection efficiency of clustering 
algorithms is reduced.
Low-density fields, characterised by a mean photon separation comparable or greater than
the PSF radius (see Sect. \ref{s:statsign}) are obtained by selecting photons with energies 
greater than a few GeV, which are characterised also by a rather stable and small PSF radius.

\begin{figure}[htb]
\centering
\smallskip
\includegraphics[width=0.45\textwidth]{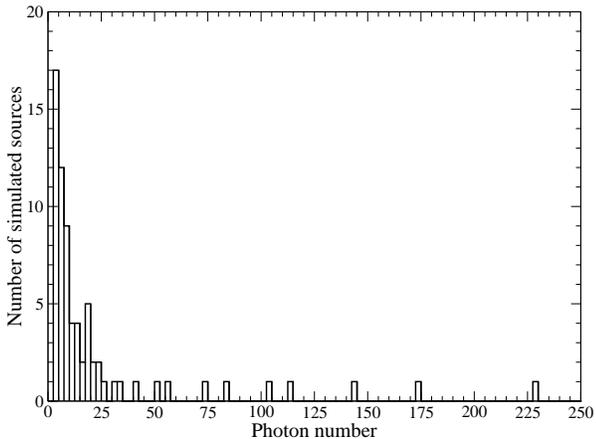}
\caption{Distribution of the number of photons in the simulated source set: note
the power law shape up to 40 photons and the uniform tail up to $\sim$250 photons.}
\label{fig:simulhisto}
\end{figure}

We considered a rather broad region covering an extension of $90^{\circ}\times25^{\circ}$ 
in Galactic coordinates, precisely, $80^{\circ} < l < 170^{\circ}$ and 
$40^{\circ} < b < 65^{\circ}$.
The 2FGL catalogue \citep{nolan12} reports 57 sources within this region: some of them 
have soft spectra and only a small number of detected photons above a few GeV.    
Using the standard tool 
\texttt{gtobssim}\footnote{\url{http://fermi.gsfc.nasa.gov/ssc/data/analysis/scitools/help/gtobssim.txt}} 
developed by the \textit{Fermi}-LAT collaboration, we simulated a 2-year long observation of the diffuse 
$\gamma$-ray background (Galactic and isotropic, using the \texttt{gll\_iem\_v02} and 
\texttt{isotropic\_iem\_v02} models) at energies above 3 GeV.
This energy, lower than that used in the previous analyses, was chosen in order to slightly increase
the photon number and to reduce possible bias in the comparison of our results with those of the
2FGL catalogue for the true LAT sky field (see Sect. 7.1).
The instrumental response function (IRF) \texttt{P6\_V3\_DIFFUSE} was used, 
combining ``front'' and ``back'' events. 
The total number of photons generated in this region is 9322.
To this photon list, we added 70 simulated sources.
For each source, the number of photons was 
chosen from a probability distribution given by a power-law, with exponent $-2$ from a minimum 
value of 4 up to 40 photons, joined to a constant tail up to 240 photons, rather similar to the 
flux distribution of the sources in 2FGL.
Photon numbers of simulated sources vary between 4 and 228, with the distribution shown in
Fig. \ref{fig:simulhisto}, for a total of 1722 photons.
Events in each source are spatially distributed with a Gaussian probability function with 
$\sigma = 0^{\circ}.2$ centered at its location.
This is a rather poor approximation of the LAT instrumental PSF at these energies, but we decided 
to use this simplified approach because in our analysis we considered only the coordinates
of individual photons and not their energy. 
The photon spatial distribution was the same for all the sources and we were able to directly compare 
it with the size of clusters found by MST.

Several simulated test fields have been generated, adding the simulated sources to the same diffuse 
background. 
The number of photons for each source is the same for all the realizations, but locations 
are randomly chosen to have different brightness contrast between sources and the surrounding
background. 

\begin{figure*}
\centering
\includegraphics[width=\textwidth]{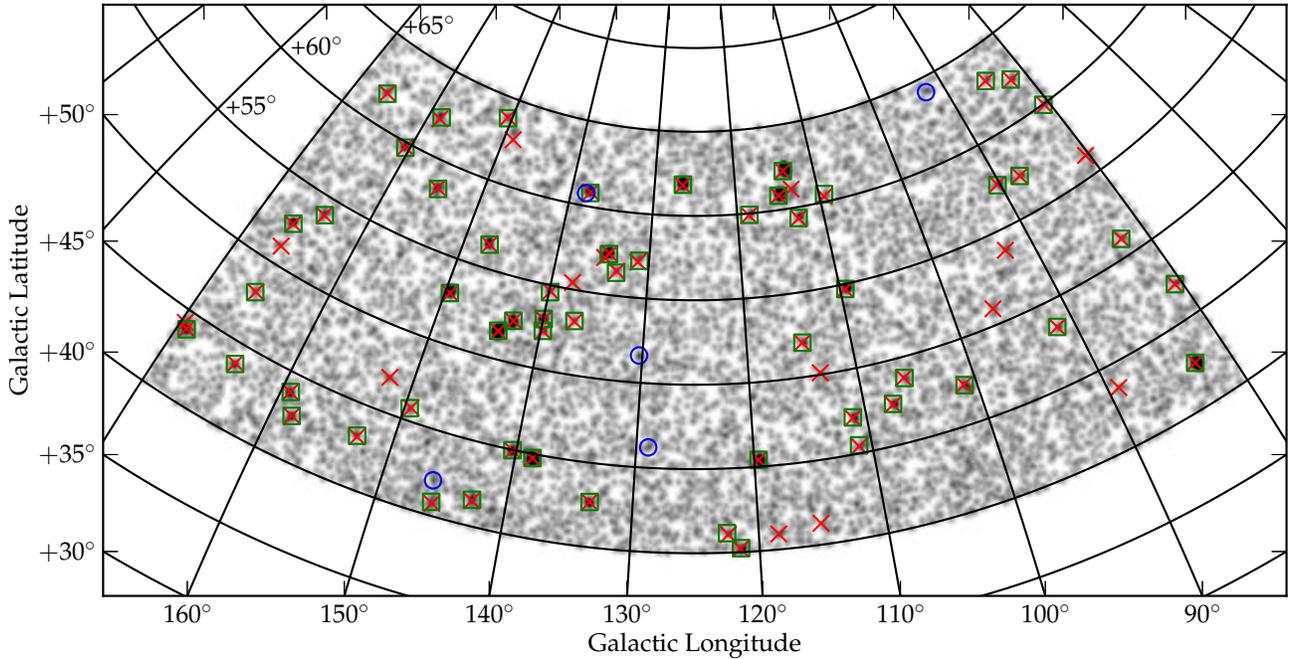}
\caption[]{Aitoff-Hammer projection in Galactic coordinates of one of the simulated 2-year test 
fields with the additional 70 sources: red crosses mark the original source positions, green
squares are the positions of clusters found by MST, blue circles are the spurious detections.}
\label{fig:camposimulato2}
\end{figure*}

\subsection{Test field results}\label{s:testfieldresults}

\subsubsection{Sensitivity}

Considering that the minimum number of photons in the simulated sources is 4, and
that some of these clusters are expected to have one or more background events in the 
close surroundings, in the primary selection we adopted $N_{\mathrm{cut}}= 3$, while 
$\Lambda_{\mathrm{cut}}$ was fixed at $0.7\,\Lambda_{\mathrm{m}}$.
This primary selection produced about 170 clusters in all the simulations.
A small increase of $\Lambda_\mathrm{cut}$ (e.g. to $0.9\,\Lambda_{\mathrm{m}}$) 
produces a higher number of clusters, because groups of nodes with larger separations between them remain connected. 
However, the majority of these clusters is characterised by low values of $g$ and photon numbers just above 
$N_{\mathrm{cut}}$, and many are usually rejected in the secondary selection. 
Therefore, there is no real advantage to use separation lengths closer to $\Lambda_{\mathrm{m}}$.
Conversely, values of $\Lambda_{\mathrm{cut}}$ equal or lower
than $0.6\,\Lambda_{\mathrm{m}}$ actually increase the appearance of satellite clusters 
and the risk of eliminating true sources with a small number of events or with low clustering
parameters.
Then the choice of a separation value equal to $0.7\,\Lambda_{\mathrm{m}}$ appears
to be a satisfactory trade-off in the practical MST application. 
MST thus finds on average $62.5 \pm3.5$ ``true'' clusters corresponding to about $90\%$ of the 70 simulated sources. 
The ``lost'' clusters are mainly composed of 4 and 5 photons: 
the average number of these is 4.7. 
This shows that the method is quite sensitive for finding clusters of more than 5 photons.
However the number of spurious clusters (i.e., of purely statistical origin) is high: 
on average $106\pm1$ spurious clusters are generated. For this reason a further selection is necessary.

\subsubsection{Spurious cluster rejection}

\begin{figure}[tbh]
\centering
\smallskip
\includegraphics[width=0.45\textwidth]{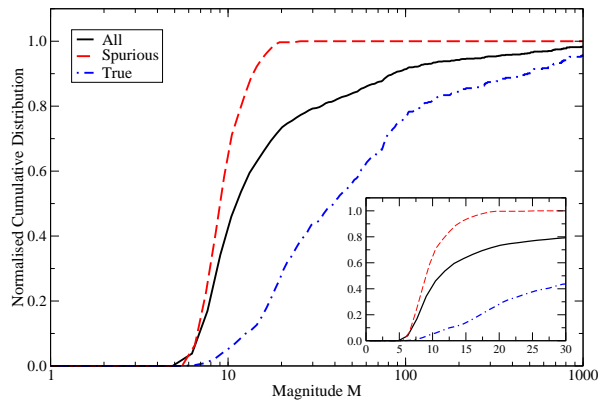}
\caption{Normalised cumulative distributions for $M$ on the simulated test fields, for all the sources (true and spurious, black continuous curve), for spurious sources (red dashed line) and true sources (blue dot-dashed line). The range $M = 0$--30 is shown in more detail and in linear coordinates in the inset.}
\label{fig:cummedio}
\end{figure}

The aim of  this secondary selection is to reach the highest rejection of clusters not
corresponding to genuine sources together with the lowest elimination of the good ones.
Many cluster parameters among those listed in Sect. 5, like $R_\mathrm{c}$, $R_\mathrm{m}$, etc., 
may be used, but for this purpose we find that $n$, $g$ and $M$ are the most efficient.
We computed for each parameter the normalised cumulative distribution (NCD) 
for spurious and true sources averaged over the simulated test fields. 
Fig. \ref{fig:cummedio} shows the NCDs for $M$, which is found to be the best parameter to reject spurious clusters with the lowest elimination of good ones. 
The red curve corresponds to the NCD of spurious clusters and the blue one to the NCD of clusters corresponding to true sources. 
$n$ is a quite good parameter also, but slightly worse than $M$ because it does not take in to account the density of photons.
To select the best value of $M$ we also computed the NCD of $M$ for all the sources (both spurious and true, black curve in Fig. \ref{fig:cummedio}).
It is easy to see that the probability $P(M_{\mathrm{cut}})$ to find true sources with $M>M_{\mathrm{cut}}$ is given by:
\begin{equation} 
P(M_{\mathrm{cut}}) = 1 - \frac{N_{\mathrm{s}}[1-C_{\mathrm{s}}(M_{\mathrm{cut}})]}{N_{\mathrm{a}}[1-C_{\mathrm{a}}(M_{\mathrm{cut}})]} \ , 
\end{equation}
where $N_{\mathrm{s}}$ and $N_{\mathrm{a}}$ are the total numbers of spurious and all sources, respectively,
and $C_{\mathrm{s}}$ and $C_{\mathrm{a}}$ are the respective NCDs.
For $M=0$ all the clusters are included and the probability is simply given by the ratio between the number of true clusters $N_{\mathrm{t}}=N_{\mathrm{a}}-N_{\mathrm{s}}$ and the total number of clusters $N_{\mathrm{a}}$. 
When the NCD of spurious clusters $C_{\mathrm{s}}$ is close to unity, the probability of finding true sources also approaches unity. 
In this way it is possible to assign a precise statistical meaning to a selection based on $M$. 
In our case, a value of $M_{\mathrm{cut}}$ around 15 gives a $90\%$ probability to find true sources.

Applying this cut, the average number of recovered true sources is reduced to $54.0\pm2.5$, 
but the average number of spurious ones is reduced to $6.7\pm0.8$.
That is, while the number of selected clusters belonging to a true source is reduced by $14\%$ upon the application of the secondary selection, the ``background'' of spurious clusters is suppressed by $94\%$. 
Observe that for $M > 20$  the probability that a cluster corresponds to a genuine source tends to unity.
However, the knowledge of $N_{\mathrm{s}}$ and $C_{\mathrm{s}}$ for real fields 
can be inferred only on the basis of simulations or statistical considerations.

In Fig. \ref{fig:camposimulato2} we reported the positions of sources (red crosses) and those of the clusters detected
by MST with $\Lambda_{\mathrm{cut}} = 0.7\,\Lambda_{\mathrm{m}}$ (green squares) for one 
simulation. 
Blue circles mark the positions of spurious clusters, i.e. having the centroid's coordinates
not matching those of the simulated sources.
Note that one of them is very close to a true source, 
and is probably a satellite cluster produced by the fragmentation of the latter.

\begin{figure}
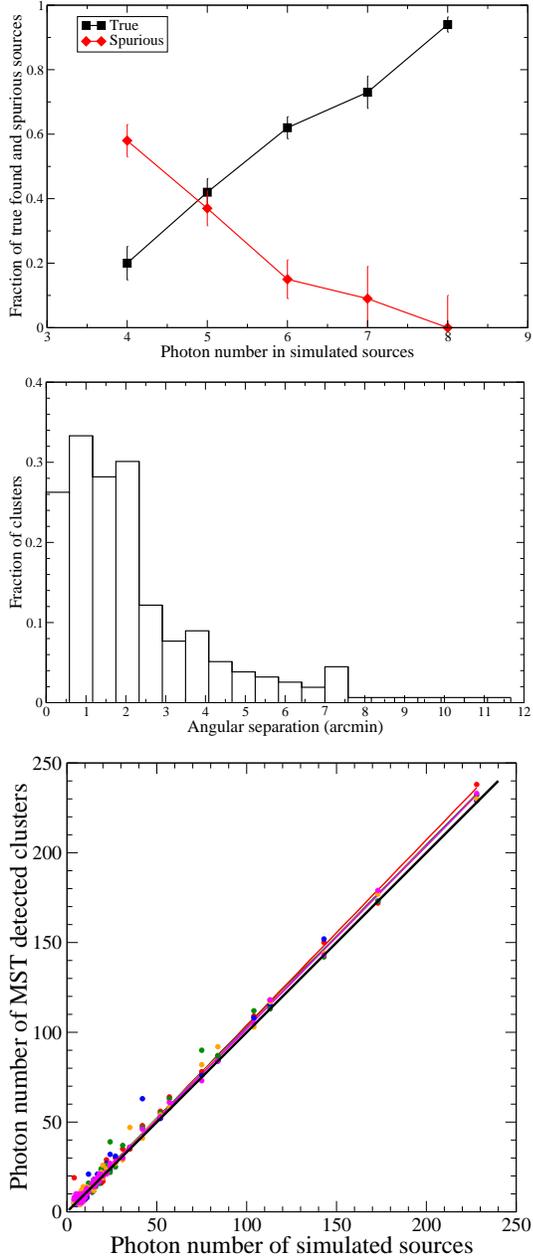

\centering
\includegraphics[width=0.4\textwidth]{fig5_a}\smallskip\
\includegraphics[width=0.4\textwidth]{fig5_b}\smallskip\
\includegraphics[width=0.4\textwidth]{fig5_c}
\caption{\emph{Upper panel:} Cumulative results for the MST analysis of the simulated test fields. The fraction of sources found or spurious is plotted, as a function of $n$.
\emph{Central panel:} Histograms of the angular distances between the source locations and the MST centroids
of associated clusters. 
\emph{Lower panel:} Scatter plot of the number of photons of clusters detected by MST against the number
of photons for some simulated test fields. The black thick solid line represents the case in which the two numbers are equal.}
\label{fig:cumulativeresults}
\end{figure}

The cumulative results of the simulations are summarized in Table \ref{tab:lostfake} and 
shown in Fig. \ref{fig:cumulativeresults}, upper panel,
which shows the mean fraction of sources found and spurious, as function of the number of 
generated photons. 

\begin{table}[htb]
\caption{MST results on the simulated test fields.
Primary selection parameters are $\Lambda_{\mathrm{cut}} = 0.7\,\Lambda_{\mathrm{m}}$ and 
$N_{\mathrm{cut}} = 3$. For the secondary selection: $M_{\mathrm{cut}} = 15$.}
\label{tab:lostfake}
\centering
\begin{tabular}{rc}
\hline
Sources found (w/o secondary select.)  & $62.5 \pm 3.5$  \\
Spurious clusters & $106 \pm 1$ \\
Sources found (w/ secondary select.) &  $54 \pm 2.5$ \\
Residual spurious clusters & $6.7\pm0.8$ \\
\hline
\end{tabular}
\end{table}

\subsubsection{Flux and positional accuracy}

The accuracy of the weighting method for the evaluation of centroid positions, described in Sect. 4,
is confirmed by the histograms of Fig. \ref{fig:cumulativeresults}, central panel, where the angular separation of the estimated positions 
from the correct ones is presented, for all the simulations together.
The large majority of sources has distances smaller than 4$'$ and only very few cases are
found more distant than 6$'$.5.
Some of the latter ones correspond to pairs of simulated sources located at very small distances, 
not separated by the adopted MST cut length and joined in a unique cluster.
Considering that the 1\,$\sigma$ radius of the Gaussian distribution used in the generation of
source photons is 12$'$ (0$^{\circ}$.2), our method for estimating the centroid position 
is confirmed to be accurate enough for a safe positional association.

Fig. \ref{fig:cumulativeresults}, lower panel, shows the scatter plot of the number of photons in the clusters detected in 
tests against the original number of simulated events.
The results are very similar in all the simulations and differences are of a few events only.
MST numbers for rich clusters are slightly in excess with respect to the original ones:
for the richest clusters the mean excess is about 2\%.
The MST method is thus able to correctly recover the right event number of true sources.
There are, however, a few cases with a rather high discrepancy. 
Again, they are related to very close pairs of sources, not resolved by the cut length.
These cases are a few percent in each simulation and can be identified only from a different
analysis.

Note that large scale inhomogeneities in the spatial distribution of photons within the entire 
field are not present (Fig.~\ref{fig:camposimulato2}) and, therefore, it was not necessary to consider smaller subregions for 
the MST analysis.

We did not find that a threshold on the median radius $R_\mathrm{m}$ and the proximity value $p$ are 
directly useful in the secondary selection, but provide more information on the extension and near 
environment of clusters.
In particular, we verified that the dispersion of $R_\mathrm{m}$ values depends on the photon number 
$n_k$: for relatively high photon numbers,  $n_k > 15$, $R_\mathrm{m}$ converges to the width of 
the Gaussian profile of simulated sources (or to the PDF effective radius in true data fields), 
while for  clusters with $n_k$ lower than this value,
it is highly dispersed and values up to $\sim$2.5 times the expected ones are found for both genuine 
or spurious clusters.
Therefore, it is not efficient in the rejection of the latter ones originating from background 
fluctuations, but can be a useful indicator for detecting anomalous extension in rich clusters possibly 
related to the occurrence of unresolved close pairs.

\section{PGW analysis and significance of the detection}\label{s:pgw}

\begin{figure*}
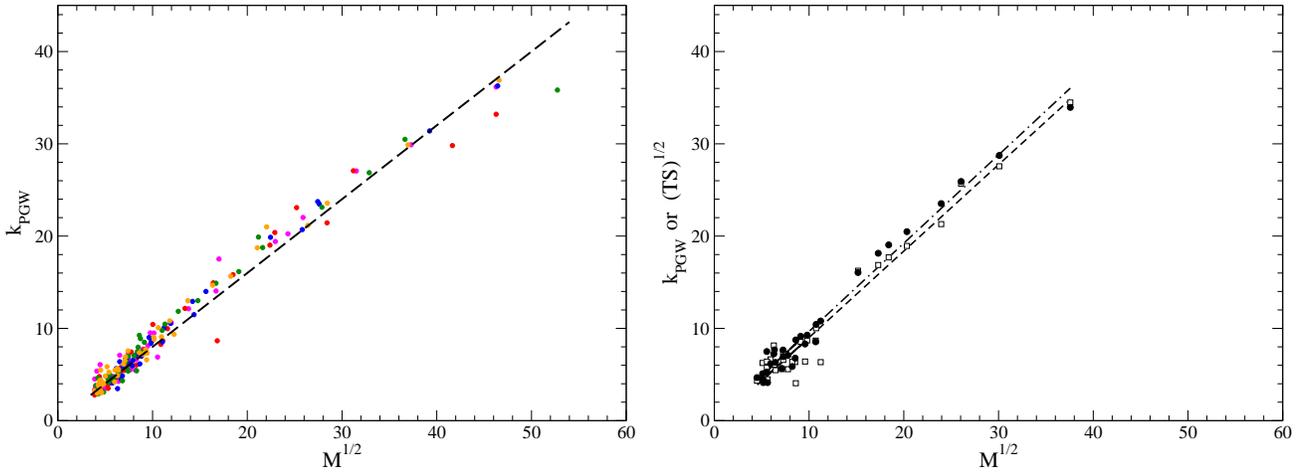

\centering
\smallskip
\includegraphics[scale=0.32]{fig6_a}\hspace{1mm}
\includegraphics[scale=0.32]{fig6_b}
\caption{\emph{Left panel}: The relation between the PGW $k$-significance and the square root of 
the magnitude $M$ in the simulated test fields. 
The thick dashed line represents the law $s = 0.8 \sqrt M$.
\emph{Right panel}: the magnitude-significance relation for the 2 year 3--10 GeV true LAT field. 
Filled circles correspond to the MST and PGW significance plot, while open squares correspond to
relation of $\sqrt M$ with $\sqrt{TS}$ for the 3--10 GeV band reported in the 2FGL catalogue.
The dashed and the dot-dashed lines represent the best fits for 
$\sqrt{M}$ vs. $k_\mathrm{PGW}$ and $\sqrt{M}$ vs. $\sqrt{TS}$ relations, respectively.
}\label{fig:SM12root}
\end{figure*}

The same simulated test fields were also analysed using the PGW
algorithm to compare its results with those obtained by MST.
This method applies a 2-dimensional Mexican-hat wavelet transform to a $\gamma$-ray image,
structured as a set of pixels whose intensity is the number of events within each of them.
Basically, it employs the algorithm described by \cite{damiani97} that
provides for each detected source, in addition to coordinates and intensity, also an 
estimate of its statistical significance measured in terms of the number of standard 
deviations above the background level, usually indicated as \emph{$k$-significance} 
(hereafter $k_\mathrm{PGW}$).
This method was already applied to simulated LAT images \citep{ciprini07} and was also used
in the preparation of the 1FGL \citep{abdo10} and 2FGL catalogues \citep{nolan12}.
PGW results confirm all the ``rich'' clusters found by MST, whereas several weak clusters
are undetected together with a small number of non-genuine sources.
Typically, about 80\% of detections are common to both methods, and MST appears to be 
slightly more efficient in the detection of genuine clusters with a small number of photons.

\subsection{Statistical significance of MST source detection}\label{s:statsign}

In \cite{campana08} we studied the probability distributions of $g$ and $n$ in uniform 
random fields generated by a Monte Carlo extraction of nodes.
A straightforward application of these results to the real sky, however, is not
practical and does not always give good estimates of the chance detection probability, 
because of the non-uniform $\gamma$-ray background and the deviation from a purely
random distribution due to the occurrence of strong sources.

We searched for a simple and reliable estimator of the statistical significance of 
individual sources by comparing the MST cluster parameters with the $k_\mathrm{PGW}$ 
values obtained by means of PGW in the simulated fields.
In principle, one can expect that the magnitude $M$ of a cluster can be such an estimator
because it depends on the number of photons and therefore can be correlated with the
PGW significance.
The plot of $\mathrm{Log}\, k_\mathrm{PGW}$ vs $\mathrm{Log}\, M$ shows, in fact, a 
very high linear correlation and a simple power law best fit gives exponents for $M$ 
close to 0.5, within a few percent.
We then assumed this value as the correct one and plotted the values of 
$k_\mathrm{PGW}$ vs $\sqrt M$ for all the simulated fields (Fig. \ref{fig:SM12root}, left panel).
The correlation between these two quantities is very high, despite a small 
number of points lying at a rather large distance from a linear trend. 
A simple relation able to describe well the data of all the simulations is 
\begin{equation}\label{eq:kpgwvsM}
k_\mathrm{PGW} =  S \sqrt M
\end{equation} 
with $S = 0.8$.
This relation can be considered as a simple practical rule for evaluating the 
statistical significance of a cluster detection by MST.
Note that, according to this criterion, the requirement that a cluster would have at least 
a significance $k_\mathrm{PGW} = 3$ would imply a threshold value for $M$ equal to 14, that 
is practically coincident to the value we found to optimize the ratio of number of lost 
sources to that of spurious ones.    

\section{Analysis of the \textit{Fermi}-LAT field }\label{s:LAT}

We verified Eq. \ref{eq:kpgwvsM} applying the two methods to the true 2-years LAT field 
used for the 2FGL catalog \citep{nolan12}, covering the same region of the simulated 
fields and selecting photons in the energy range 3--100 GeV and detected between 2008 August 5 and 2010 August 5.
This field is well suited for a cluster analysis because the density gradient of photons
toward low Galactic latitudes is mild: the mean angular separation between photon pairs
changes from about 0$^\circ$.25 close to the high $b$ boundary to about 0$^\circ$.2 close 
the lower one.
It is therefore always comparable with the radius of the instrumental PSF in this energy
range.
LAT events were selected in a rectangular box corresponding to the test field Galactic coordinates, 
and we applied the standard cut on the zenith angle ($<$100$^\circ$) and on the rocking angle ($<$52$^\circ$) to limit contamination from Earth limb $\gamma$-rays.
The 2FGL catalogue reports 57 sources  with $TS > 25$ in the 100 MeV--100 GeV 
energy range inside the true LAT test field.
Only 38 of these are detected above 3 GeV, and 12 of this latter group have $TS$ lower than 25 in this
energy range.
The application of MST with the criteria for secondary selection given above ($M>15$) finds 
39 significant clusters associated with the 2FGL sources: 37 correspond to 2FGL sources
detected above 3 GeV and the remaining 2 correspond to 2FGL sources detected only in lower energy bands; 
the only 2FGL source with a measured flux in the 3--10 GeV energy band and not detected by MST has the quite low $TS=9.6$ in this band.

Among the sources detected by the two methods, we selected a subset of 32 sources having 
$k_\mathrm{PGW} >4$ and plotted their $k$-significance vs $\sqrt M$ (Fig. \ref{fig:SM12root}, right panel): 
again a high linear correlation resulted.
A similar correlation is also found with the ML significance $\sqrt{TS}$
in the 3--10 GeV band, extracted from the 2FGL catalogue.
In both cases, the best fit values of $S$ were very close to unity: 0.94 and 0.96,
respectively. 
Note that the residuals of data points with respect to best fit lines are higher for the 
2FGL $\sqrt{TS}$ set than for $k_\mathrm{PGW}$, particularly for sources with 
$\sqrt M < 15$ and, therefore, with signal-to-noise ratios that are not very high.
This effect can be due to the fact that $\sqrt{TS}$ in the catalogue were evaluated in an 
energy band narrower than the one used by us, and therefore photon numbers of some individual
sources can be lesser than ours.  

An intuitive explanation of Eq.~\ref{eq:kpgwvsM} can be based on the quasi-Poissonian 
spatial distribution of photons in the field.
In this case, in fact, one can reasonably associate with a cluster of $n_k$ photons, 
(which is one of the two factors in the definition of $M$) a standard deviation equal 
to $\sqrt n_k$.
Considering that the range of the clustering degree $g$ is generally limited from 
2.5 to $\sim$10, with the large majority of values lower than 6 (in our simulations 
the mean values of $g$ were always close to 4.2 with a rms value of 1.5), the additional 
variance due to this quantity is not large.
Thus a linear relation between $k_\mathrm{PGW}$ and $\sqrt M$ would be expected.
We verified in our simulations also the occurrence of a high linear correlation between 
$\sqrt n_k$ and $k_\mathrm{PGW}$. 
Nevertheless the use of the magnitude, instead of the number of photons, appears 
more convenient for the following reasons: 
$i)$ the secondary selection based on $M$, that takes implicitly into account the cluster 
density, is more efficient in selecting the poorest clusters than a procedure based on $n_k$;
$ii)$ the parameters of the linear regression are found more stable when the different 
fitting intervals in $\sqrt n_k$ or $\sqrt M$ are considered.  
We do not have a complete theoretical explanation for this fitting stability and for computing
the expected value of $S$. 
Anyway, our results suggest the following practical approach: 
having obtained a list of common clusters with MST and another method, one can select a sample 
of high significance sources in the latter set to be used for evaluating the best $S$ value.

\section{New candidate $\gamma$-ray sources}\label{s:newsources}

In addition to clusters associated with already known $\gamma$-ray sources, our MST 
analysis of the LAT test field provided a number of clusters having rather high values 
of $M$, and therefore a high confidence to be not spurious.
Nine clusters with $M > 19$, 6 of them with $n > 8$, are reported in Table 
\ref{tab:testfieldsrc};
these threshold values did not give spurious detections in the simulated field analysis (Sect. \ref{s:testfieldresults}).
They are divided into two groups, based on their number $n$ of photons ($n\ge9$ or $n<9$), the former ones are expected to have a higher confidence 
to be genuine. 

We also searched for possible counterparts to be associated with these clusters to make their detections 
more robust as genuine $\gamma$-ray emitters. We found interesting positional
correspondences within an angular distance of 0$^\circ$.2 for four of them with BL Lac objects 
from the {\it Roma-BZCAT} \citep{massaro09b, massaro11}.
Considering that this catalogue reports 115 BL Lac objects and candidates within the 
entire considered LAT field, which covers about 1360 square degrees, their mean 
density is about 0.085 sources~deg$^{-2}$.
The resulting chance probability to find one BL Lac object within a circle with a radius of 
0$^\circ$.2 is $\sim$1.1$\times10^{-2}$.
Thus, we obtain a chance probability for 4 associations 
out of these 9 trials
of the order of 3$\times10^{-6}$.
It appears very unlikely that these associations would be found if all clusters were spurious.
Note that we did not find possible counterparts classified as flat spectrum radio quasars, in
agreement with the expectation that they generally have soft $\gamma$-ray spectra and
therefore a low photon flux above 3 GeV \citep{ackermann11}.
The possibility to find radio sources within the same angular radius is, of course, much
higher, but the majority of them are generally rather weak and without optical counterparts. 
Only in two cases we found bright radio sources (also reported in Table \ref{tab:testfieldsrc}) 
but they do not exhibit definitive blazar characteristics and their association with $\gamma$-ray 
emission appears unlikely.

\begin{figure*}
\centering
\includegraphics[scale=0.5]{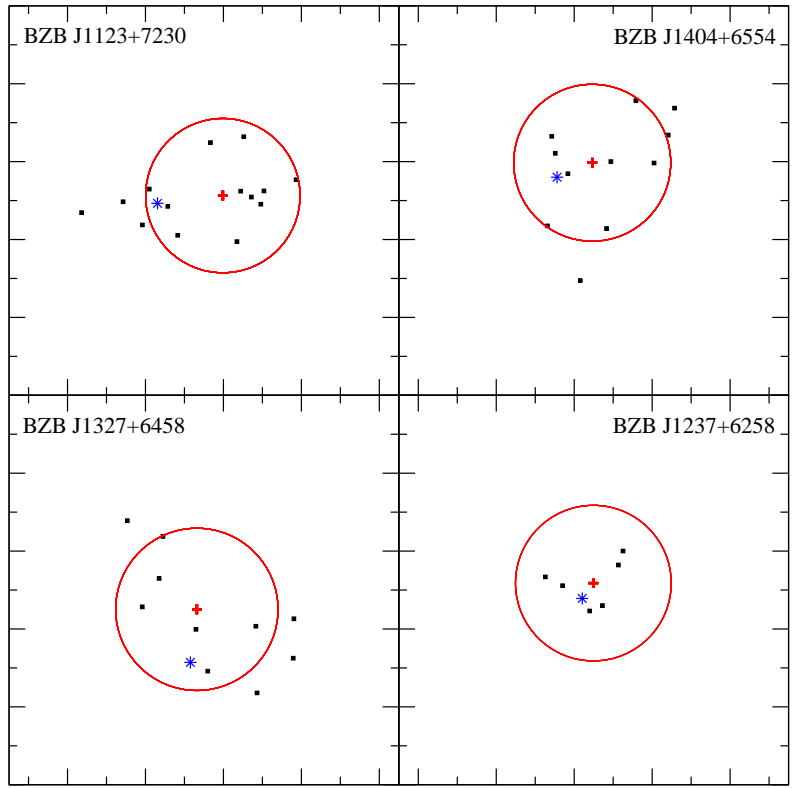}
\caption{Photon maps (between 3 and 100~GeV) of four new detected clusters having possible BL Lac counterparts. 
Each panel is $1^{\circ} \times 1^{\circ}$ wide and oriented with the Galactic coordinates.
Photon longitude distances from the centroid have been scaled by the cosine of the latitude.
Red crosses mark the centroid positions and the circles delimit regions having a radius
of 0$^\circ$.2.
Blue stars mark the coordinates of possible BL Lac counterparts: BZB~J1123+7230 (top-left),
BZB~J1327+6458 (bottom-left), BZB~J1404+6554 (top-right), BZB~J1237+6258 (bottom-right).
}\label{fig:newclustmap}
\end{figure*}

A further interesting finding is that four clusters with $n > 8$ have $g$ values lower than 2.5, 
and two of them slightly lower than 2, indicating that their ``clumpiness'' is low.
Among the 37 clusters associated with 2FGL, 18 have $g < 3$, and 4 have 
$g < 2.5$.
Some of the newly detected features could be associated with extended structures or with pairs of 
sources having a small angular separation, comparable to the instrumental resolution, 
and therefore they can be missed by algorithms based on the matching with the PSF shape.
A more detailed analysis over regions having a size of 10$^\circ$ and centered at the 
cluster likely associated with BZB~J1123+7230 separates it into two smaller groups: one 
having 6 photons ($M = 11.99$) closer to the blazar position, and the other with 8 photons 
($M = 22.83$) and having coordinates RA = $170^\circ.48$, Dec = $72^\circ.41$ 
without a known possible counterpart.
Note also that the latter cluster is found at energies higher than 6 GeV, with 6 photons
and $M = 18.2$.
Applying the shorter separation length $\Lambda_{\mathrm{cut}} = 0.5\,\Lambda_{\mathrm{m}}$ 
the two clusters are split again, one of them with $n = 5$ and $M = 24.0$. 
A similar analysis for the cluster close to BZB~J1327+6458, which has a rather low $g$ 
value, gives a detection above 6 GeV with 7 photons and $M = 16.8$, but a centroid position 
slightly more distant from the possible BZB counterpart, and about at the same separation 
from the radio source NVSS~J074638+520042, of which no optical counterpart is known. 
We cannot exclude, therefore, the occurrence either of a possible confusion or of a spurious
association.
The other two associations with BZB sources reported in Table 3, appear to be positionally 
more robust, in particular the one with BZB~J1237+6258, that has a high $g$ although with 
only 6 photons.
This source is located very close to the southern boundary of the field, a circumstance
that can affect the construction of the MST; however, a test for the existence of this cluster
in a surronding region 10$^\circ$ size around it confirmed its detection.
Again, the analysis of the cluster associated with BZB~J1404+6554 at energies above 6 GeV 
confirms a detection with 7 photons and $g=2.82$.
Finally, we note that only the cluster at  RA~=~$207^\circ.32$, Dec~=~$71^\circ.53$,
for which no possible counterpart has been found, does not survive the more severe cut. 
Photon maps of these four clusters are given in Fig. \ref{fig:newclustmap}.

We also searched for these clusters with the PGW algorithm and the standard ML 
\citep[ML][]{abdo10} analysis. 
The former method finds only three detections with a significance 
in the range between 2.5 and 3 standard deviations. 
In Table \ref{tab:testfieldsrc},  the value of the Test Statistic ($TS$) for each new MST source candidate was derived applying the binned likelihood method implemented in the standard LAT Science Tools\footnote{\url{http://fermi.gsfc.nasa.gov/ssc/data/analysis/scitools/overview.html}} (version v9r28p0). 
We analyzed data collected, in the 3 GeV--300 GeV energy range, from 2008 August 5 to 2011 August 5.  For this analysis, only events belonging to the Pass 7 ``Source'' class and located in a circular region of interest (ROI) of 10$^\circ$ radius, centered at the position of the MST candidate source were selected. We applied the standard cuts on the zenith angle ($<$100$^\circ$) and on the rocking angle ($<$52$^\circ$).
The ROI model used by the binned likelihood method includes all point sources from the 2FGL catalog \citep{nolan12} located within 15$^\circ$ of the MST candidate source. Sources located within a 2$^\circ$  radius centered on the MST source position had all spectral parameters left free to vary during the fitting. The MST candidate was modeled as a power law with both normalisation and spectral index free to vary. We used IRF version \texttt{P7SOURCE\_V6}\footnote{\url{http://www.slac.stanford.edu/exp/glast/groups/canda/lat_Performance.htm}}. The diffuse Galactic and isotropic components were modeled with the same ones used for generation of 2FGL (\texttt{gal\_2yearp7v6\_v0} and \texttt{iso\_p7v6source} files\footnote{\url{http://fermi.gsfc.nasa.gov/ssc/data/access/lat/BackgroundModels.html}}). The normalisations of the components comprising the total background model were allowed to vary freely. 
For all the MST candidate sources reported in Table \ref{tab:testfieldsrc} we found  $TS < 25$ (with 2 degrees of freedom, spectral index and normalisation, $TS=25$ roughly corresponds to a 4.6$\sigma$ detection significance).

\begin{table*}
\caption{Coordinates and main properties of MST clusters detected in the 2 year LAT test field 
and not associated with sources in the 2FGL catalogues. Maximum likelihood significance is reported
only when $TS > 9$. Celestial coordinates are J2000.}
\centering
{\small
\begin{tabular}{rrrrrccrlll}
\hline
RA~~~ & Dec ~~~& $l$~~~ & $b$~~~ & $n$ & $g$~~~ & $M$~~~ & PGW & $TS$ & ang. dist. & Possible \\
 deg~~ &  deg~~ & deg~~ & deg~~ & &  &  &  det.&  & arcmin & counterparts\\
\hline
170.49 & 72.40 & 132.00 & 43.11 & 14 & 2.652 & 37.128 & yes & 24.3 & 10 & BZB J1123+7230 ($^1$)\\
179.00 & 61.61 & 134.18 & 54.29 &  9 & 2.427 & 21.843 &  no & 11.6 &       &           ---    \\ 
201.89 & 64.84 & 116.73 & 51.85 & 10 & 1.941 & 19.410 & yes &    ---  &  8 & BZB J1327+6458 ($^1$)\\
207.32 & 71.53 & 116.52 & 44.87 & 10 & 1.956 & 19.560 & no  &   ---   &       &       ---        \\ 
210.97 & 65.91 & 111.65 & 49.60 & 11 & 2.141 & 23.551 & yes &    ---  &  6 & BZB J1404+6554; 4C 66.14 ($^2$) \\
237.64 & 60.34 &  93.32 & 45.14 &  9 & 2.746 & 24.714 &  no & 15.6 &    ---   &       ---        \\ 
\hline
142.46 & 66.38 & 146.76 & 40.17 &  8 & 2.408 & 19.264 & no & ---     & 6 & TXS 0925+665 ($^3$)       \\ 
158.13 & 69.19 & 139.08 & 43.32 &  7 & 2.998 & 20.986 & no & ---     & ---     &      ---         \\ 
189.36 & 62.94 & 125.65 & 54.12 &  6 & 3.338 & 20.028 & no & 12.6 & 3 & BZB J1237+6258 \\  
\hline
\end{tabular}
\flushleft{1: possible pair of clusters.}\\
\flushleft{2: bright compact and steep spectrum source at a distance of 3$'$.47 from the cluster centroid. }\\
\flushleft{3: steep spectrum radio source with a bright core and two possible lobes of much weaker brightness.}\\
}
\label{tab:testfieldsrc}
\end{table*}

\section{Conclusions}\label{s:conclusions}

The application of MST to true and simulated LAT images confirms the validity of this method 
for searching $\gamma$-ray sources, especially for weak ones.

We defined and investigated the properties of the parameter $M$, or magnitude, defined as the 
product of the clustering degree and the number of nodes in a given tree, that was found to 
be very useful in the selection of statistically robust clusters.
An application to simulated and true LAT fields, in comparison with the PGW method, has shown
that $\sqrt M$ gives also a quite good estimate of the statistical significance of a source.
Therefore, a secondary selection based simply on a threshold on $M$ close to 15 would reject the 
large majority of spurious (low significance) clusters.
A few other spurious sources can be eliminated by further selections based on the clustering
degree $g$.
For a more complete definition of source candidates, it is useful to consider other parameters. 
The proximity value, i.e. the angular separation to the closest cluster, helps in extracting 
possible satellite clusters surrounding intense sources and possible extended structures.
The cluster and median radii can also provide additional information on that, in particular 
for the richest clusters.

Our simulated photon fields --- satisfying \emph{only} the conditions to have a rather uniform density with average photon angular distances comparable with the LAT PSF --- show that sources with a number of photons higher than 6 are expected to have a very high probability to be detected.

The probability of finding a spurious cluster, i.e. not corresponding to a genuine source, 
increases for smaller numbers of events, but for $4\leq n \leq 6$ events this probability is of 
the order of 15\%--58\% (Figure \ref{fig:cumulativeresults}).
Therefore, one can reasonably expect that some of them could be found to be 
actually associated with counterparts at different frequencies. 
On the other hand, for $n\geq7$ the probability to find a spurious cluster is lower than 10\%. 

The application of MST --- with the criteria for secondary selection given above --- to the true LAT test field
shows that the method is rather sensitive to find sources that have quite low statistical significance (TS) according to the ML analysis.
Moreover MST detects some clusters that appear to be statistically significant according to the method itself, but show a low or no significance according to other methods (PGW and ML).
Some of these clusters show plausible associations with possible counterparts, mainly with BL Lac objects.

The particular advantage in using MST for $\gamma$-ray source detection is that its cluster selection is based on the requirement that photon separations must be shorter than a given threshold, and does not directly depend on the density contrast over sampling size regions. 
It is therefore well suited to find clusters with a low photon number but high density.
Moreover, because it works directly on photon coordinates and does not use pixellated images, it is independent of the actual PSF shape and is able to detect paired clusters that could be missed by other methods.

The limitation of our MST algorithm is that it appears, at the moment, well suited to analyze fields without large gradients in the photon distribution and where the mean photon separation is comparable to the PSF radius. Source detection in fields with a high density of photons, in which the mean angular distance between them is much less than that the instrumental PSF size, and also in fields with large inhomogeneities on different angular scales, is generally a thorny problem for non-local algorithms. In these conditions one can expect that the number of spurious clusters and the fragmentation of strong sources in several clusters make the proximity analysis difficult. 
For long separation distances, in several cases extended structures not compatible with a point source are selected. Improved selection criteria should then be developed in this case, possibly to be adapted for the search of particular classes of $\gamma$-ray sources.

In conclusion, our analysis gives strong indications that the MST algorithm is a very good tool to identify small clusters, corresponding to faint sources (which can be associated with sources of low brightness, thus gaining more information on the population of low luminosity blazars and other AGNs).
In general, the clusters selected by MST should be further analyzed with other methods able to evaluate the flux and the spectrum of the sources.
MST, also considering its limitations and advantages, is able to
quickly provide a list of seed sources to be tested with ML, 
that can contain additional information (e.g. on cluster compactness, as given by $g$ and $M$) 
with respect to other cluster finding algorithms.

It should be emphasized that the computation of the MST directly provides information
on the mean angular distance between the nodes in the individual 
clusters and its ratio to that of the field.
Consequently, the $M$-based selection can be easily applied to
extract clusters having a high probability to be genuine sources
without performing other assessments on the local background 
properties.

The computational time issue is unimportant for our applications to source detection in gamma-ray images.
On the basis of our practical work, typical computation time in fields containing a number
of photons of the order of 50,000 or more is actually fast enough\footnote{As an example, for a field containing 50,000 photons, the computational time is shorter than 20 seconds on a 2.2 GHz Intel Core i5 laptop with 4 GB RAM. The secondary selection algorithms account for more than 90\% of the overall running time.}g
and even if it would be longer, this will not be a problem. 
MST analysis, in fact, is not run in real time during the observations, where a fast result
can be necessary, but on off-line archival data accumulated in many months or years of observations.
The secondary selection analysis is more time consuming. However, this direct study 
of the cluster structure is useful to understand the nature of peculiar clusters, such those 
having a large number $N$ of photons but a low clustering degree.

As a final note, we would like to stress that astrophysical applications of our developments of 
the MST method are not limited to $\gamma$-ray astronomy, but can be easily extended to other 
research fields where the problem is to search for clustered structures against a background with 
a random spatial distribution, like in the case of finding star or galaxy groups and clusters or 
other clumped structures.

\section*{Acknowledgments}
We are grateful to the referee, Maria Concetta Maccarone, whose helpful comments greatly 
improved the quality of the text, and to Toby Burnett, Seth Digel and Andrea Tramacere for useful discussions.
This research has been partially supported by a grant from Universit\`a di Roma
``La Sapienza''.
The \textit{Fermi}-LAT Collaboration acknowledges generous ongoing support
from a number of agencies and institutes that have supported both the
development and the operation of the LAT as well as scientific data analysis.
These include the National Aeronautics and Space Administration and the
Department of Energy in the United States, the Commissariat \`a l'Energie Atomique
and the Centre National de la Recherche Scientifique/Institut National de Physique
Nucl\'eaire et de Physique des Particules in France, the Agenzia Spaziale Italiana
and the Istituto Nazionale di Fisica Nucleare in Italy, the Ministry of Education,
Culture, Sports, Science and Technology (MEXT), High Energy Accelerator Research
Organization (KEK) and Japan Aerospace Exploration Agency (JAXA) in Japan, and
the K.~A.~Wallenberg Foundation, the Swedish Research Council and the
Swedish National Space Board in Sweden.
Additional support for science analysis during the operations phase is gratefully 
acknowledged from the Istituto Nazionale di Astrofisica in Italy and the 
Centre National d'\'{E}tudes Spatiales in France.

\bibliographystyle{spr-mp-nameyear-cnd}
\bibliography{bibliography} 
\label{lastpage}
\end{document}